\documentclass[journal]{IEEEtran}
\usepackage{amssymb,amsmath}
\usepackage{amsthm}
\usepackage{graphicx}  
\usepackage{hyperref}
\usepackage{multirow}
\usepackage{stmaryrd}
\usepackage{algorithm}
\usepackage{algpseudocode}
\usepackage{graphicx}
\usepackage{caption}
\usepackage{subcaption}\usepackage{hyperref}
\usepackage{color,soul}
\newtheorem{definition}{Definition}

\newenvironment{example}{\begin{itshape}%
 {\scshape Example. }%
}{%

	\end{itshape}
}

\usepackage{enumitem}

\algtext*{EndWhile}
\algtext*{EndIf}
\algtext*{EndFor}
\algtext*{EndFunction}
\begin{document}
\title{Norm Monitoring under Partial Action Observability}
\author{Natalia Criado
        and Jose M. Such,~\IEEEmembership{Member, ̃IEEE}
\thanks{N. Criado is with the School of Computing and Mathematical Sciences, Liverpool John Moores University, UK, e-mail: n.criado@ljmu.ac.uk.}
\thanks{J. M. Such is with the School of Computing and Communications
Infolab21, Lancaster University, UK, email:j.such@lancaster.ac.uk.}
}
\maketitle
\setlength{\tabcolsep}{1.5pt}
\begin{abstract}
In the context of using norms for controlling multi-agent systems, a vitally important question that has not yet been addressed in the literature is the development of mechanisms for monitoring norm compliance under partial action observability. This paper proposes the reconstruction of unobserved actions to tackle this problem. In particular, we formalise the problem of reconstructing unobserved actions, and propose an information model and algorithms for monitoring norms under partial action observability using two different processes for reconstructing unobserved actions. Our evaluation shows that reconstructing unobserved actions increases significantly the number of norm violations and fulfilments detected.
\end{abstract}

\IEEEpeerreviewmaketitle
\section{Introduction}
Within the Multi-agent System (MAS) area, norms are understood as means to coordinate and regulate the activity of autonomous agents interacting in a given social context \cite{y2006normative}. The existence of autonomous agents that are capable of violating norms entails the development of norm control mechanisms that implement norms in agent societies. 

In the existing literature, several authors have proposed infrastructures to observe agent actions and detect norm violations upon them \cite{DBLP:conf/ijcai/AlechinaDL13,DBLP:journals/eaai/MeneguzziMOMLF12}. The majority of these proposals have focused on providing efficient and scalable methods to monitor norms in dynamic agent societies, but they assume that all actions of agents are observable. However, this assumption is too strong because it is not necessarily true that all actions to be controlled can always be observed. One reason for this is that observing actions usually entails high costs. For example, the costs of setting, maintaining, and managing traffics radars to detect car speeds are very high, so traffic authorities usually decide to install a few of them in specific and critical locations. Another reason is that illegal actions may take place outside the institution controlled by the monitor; however, the effects of these actions can still be detected within the institution. For example, black market transactions cannot be directly observed by legal authorities, yet the corresponding money laundering transactions can be detected and sanctioned by these authorities. 

Very recent work on norm monitoring under partial action observability proposes solutions to ensure complete action observability by increasing the actions that are observed, either by adding more monitors \cite{bulling2013monitoring} or by adapting the norms to what can be observed \cite{alechina2014norm}. However, these solutions are not always appropriate or feasible. For instance, in e-markets, such as eBAY\footnote{\url{http://www.ebay.com}} or Amazon\footnote{\url{http://www.amazon.com}}, it is not possible to change trading laws to what can be observed. This paper goes beyond these approaches by also considering actions that were not observed but that can be reconstructed from what was observed.

The main contributions of this paper are: (i) a formalisation of the problem of reconstructing unobserved actions from observed actions for the purpose of norm monitoring; (ii) an exhaustive and an approximation solution to this problem; and (iii) an information model and algorithms used to monitor norms under partial action observability. Through an extensive empirical evaluation, we show that reconstructing unobserved actions increases noticeably the number of norm violations and fulfilments detected.

This paper is organised as follows: Section \ref{sec:def} contains the preliminary definitions used in this paper. Section \ref{sec:NM} describes the information model of norm monitor proposed in this paper. Section \ref{sec:Alg} contains the algorithms executed by norm monitors. Our proposal is evaluated in Section\ref{sec:eva}. Related word is discussed in Section \ref{sec:rw}. Finally, conclusions are contained in Section \ref{sec:con}.

\section{Preliminary Definitions}\label{sec:def}

$\mathcal{L}$ is a first-order language containing a finite set of predicate and constant symbols, the logical connective $\lnot$, the equality (inequality) symbol $=$ ($\neq$), the true ($\top$) and false propositions
($\bot$), and an infinite set of variables. The predicate and constant symbols are written as any sequence of alphanumeric characters beginning with a lower case letter. Variables are written as any sequence of alphanumeric characters beginning with a capital letter. We also assume the standard notion of substitution of variables \cite{fitting1996first}; i.e., a substitution $\sigma$ is a finite and possibly empty set of pairs $Y/y$ where $Y$ is a variable and $y$ is a term.

The set of grounded atomic formulas of $\mathcal{L}$ is built of a finite set of predicates and objects that characterise the properties of the world relevant to norm monitoring. By a situation, we mean the properties that are true at a particular moment. Some of these properties are static and not altered by action execution, whereas other properties are dynamic and changed due to agent actions. Specifically, we represent static properties as a set\footnote{In this paper sets are to be interpreted as the conjunction of their elements.} of atomic grounded formulas of $\mathcal{L}$, denoted by $g$. A state $s$ is a set of grounded atomic formulas of $\mathcal{L}$, describing dynamic properties which hold on state $s$. Thus, a situation is built on a ``closed assumption'' and defined by a set of static properties $g$ and a state $s$. Moreover, there is a set of inference rules ($\triangledown$) representing domain knowledge. 

\begin{figure}
	\centering
        \begin{subfigure}[b]{0.15\textwidth}
                \includegraphics[trim = 0mm 220mm 85mm 0mm, clip, width=\textwidth,page=1]{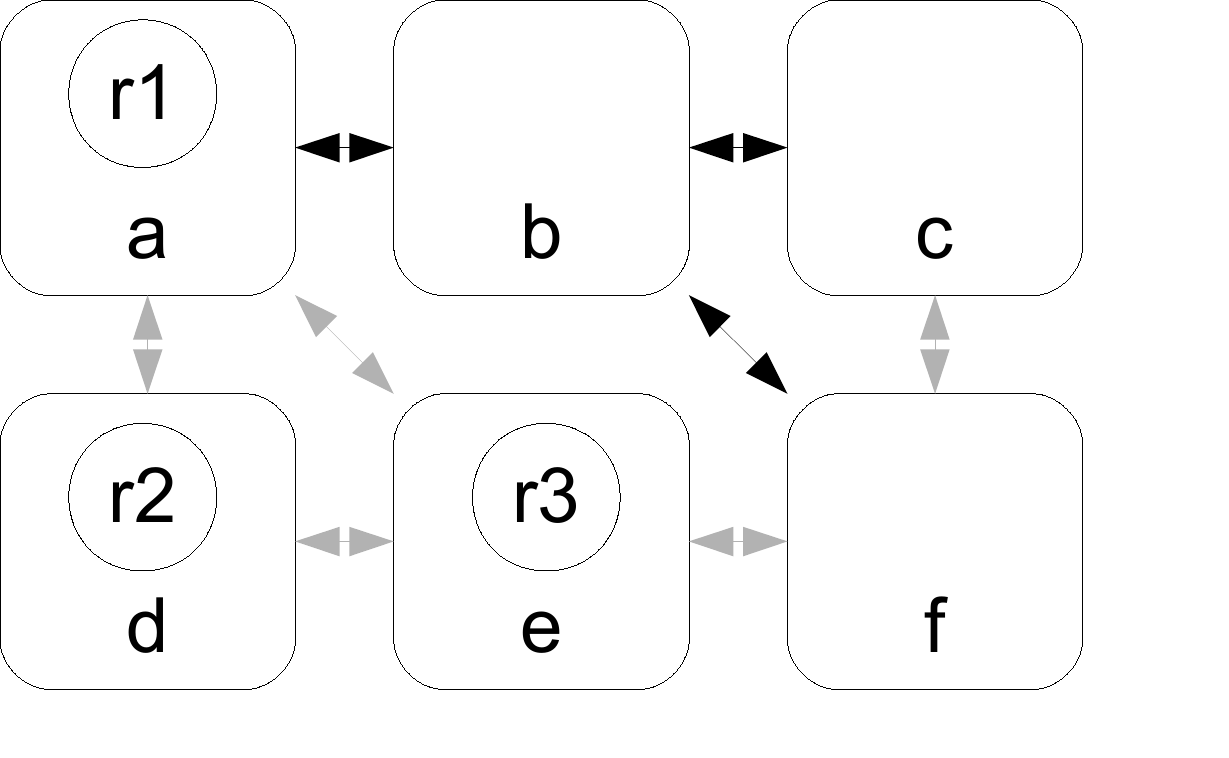}

                \caption{Initial State $s_{0}$}
                \label{fig:initial}
        \end{subfigure}%
        ~ 
        \begin{subfigure}[b]{0.15\textwidth}
                \includegraphics[trim = 0mm 220mm 85mm 0mm, clip, width=\textwidth,page=2]{Example.pdf}

                \caption{State $s_{1}$}
                \label{fig:first}
        \end{subfigure}
        ~ 
        \begin{subfigure}[b]{0.15\textwidth}
                \includegraphics[trim = 0mm 220mm 85mm 0mm, clip, width=\textwidth,page=3]{Example.pdf}
                \caption{State $s_{2}$}
                \label{fig:second}
        \end{subfigure}

	\caption{Example Scenario. Offices are represented by squares, agents are  represented by circles and the corridors are represented by arrows. Black arrows correspond to corridors observed by the Norm Monitor (NM) and grey arrows correspond to corridors not observed by the NM.}
	\label{fig:ExampleScenario}
\end{figure}

\begin{example}
In this paper we will use a running example in which there are three robots that should attend requests at six offices in a building. The goal of the robots is to attend these requests as soon as possible. Figure \ref{fig:initial} depicts our initial scenario. In our example, the language $\mathcal{L}$ contains: 4 predicate symbols ($robot,\mathit{office},in,corridor$), used to represent the robots and offices, the positions of the robots and the connections between offices in the building; 3 constant symbols to represent the robots ($r1,r2,r3$); and 6 constant symbols to represent the offices ($a,b,c,d,e,f$). The information about the robots, offices and corridors between offices is static and represented as follows:
\begin{align*}
 g = \{& robot(r1), robot(r2),robot(r3),\mathit{office}(a),...,\mathit{office}(f),\\
 &corridor(a,b),corridor(b,a),..., corridor(e,a)\}
\end{align*}
The information about the location of the robots is dynamic. Specifically, the initial state $s_{0}$ is defined as follows:
\[s_{0}= \{in(r1,a),in(r2,d),in(r3,e)\}\]
In this domain there is an inference rule ($\triangledown$) representing that a robot cannot be in two different offices at the same time: 
\[\triangledown=\{\{in(R1,OA), in(R1,OB), OA\neq OB\}\vdash \bot\}\]
\end{example}

\subsection{Action Definitions}
$D$ is a finite set of action descriptions that induce state transitions. An action description $d$ is represented using preconditions and postconditions. If a situation does not satisfy the preconditions, then the action cannot be applied in this situation. In contrast, if the preconditions are satisfied, then the action can be applied transforming the current state into a new state in which all negative literals appearing in the postconditions are deleted and all positive literals in the postconditions are added. Moreover, actions are executed in a MAS and, as a consequence, we need to be able to represent concurrent actions with interacting effects. For the sake of simplicity, we will represent concurrent actions without an explicit representation of time\footnote{An explicit representation of time may play a role on other problems like scheduling concurrent actions, but is not strictly necessary for monitoring the effects of interaction.} as proposed in \cite{boutilier2001partial}. The main idea beyond this representation is that individual agent actions do interact (i.e., one action might only achieve the intended effect if another action is executed concurrently). Specifically, each action is also represented by a (possibly empty) concurrent condition that describes the actions that must (or cannot) be executed concurrently\footnote{A more sophisticated definition of the concurrent condition would allow actions to have conditional effects according to the actions that are executed concurrently. Without loss of expressiveness, we will not consider conditional effects in action descriptions (note that any action with conditional effects can be represented by a set of actions with non conditional effects).}. 

\begin{definition}
An \textit{action description} $d$ is a tuple $\langle name, \allowbreak pre,con,post\rangle$ where:
\begin{itemize}
	\item $name$ is the action name;
	\item $pre$ is the precondition, i.e., a set of positive and negative literals of $\mathcal{L}$ (containing both dynamic and static properties) as well as equality and inequality constraints on the variables;
	\item $con$ is the concurrent condition; i.e., a set of positive and negative action schemata\footnote{An action schema contains an action name and the parameters of this action. Note that positive action schemata are implicitly existentially quantified --i.e., one instance of each positive schema must occur concurrently-- and negative schemata are implicitly universally quantified.}, some of which can be partially instantiated or constrained;
	\item $post$ is the postcondition; i.e., a set of positive and negative literals of $\mathcal{L}$ (containing dynamic properties only).
\end{itemize}
\end{definition}
Given an action description $d$, we denote by $pre(d), con(d), \allowbreak post(d)$ the action precondition, concurrent condition and postcondition.

\begin{example}
In our example, there is only one action that can be executed by robots:
\[\left\langle
\begin{array}{l}
	move,\{robot(R),\mathit{office}(O1),\mathit{office}(O2),in(R,O1),\\
	corridor(O1,O2)\},\{\},\{\neg in(R,O1),in(R,O2)\}
  \end{array}
\right\rangle\]
This action represents the movement of a robot from one office to another. The parameters of this action are the robot ($R$), the source office ($O1$), the destination office ($O2$). To execute this action, the robot should be located at the source office and the two offices should be connected. Once the operation has been applied, the robot is no longer at the source office and it is at the destination office. 
\end{example}

\begin{definition}
Given a situation represented by the state $s$ and a set of static properties $g$, and an action description $d=\langle name, pre,con,post\rangle$; an \textit{action instance} (or action) is a tuple $\langle name, pre',\allowbreak con',post'\rangle$ such that:
\begin{itemize}
	\item There is a substitution $\sigma$ of variables in $pre$, such that the precondition is satisfied (i.e., entalied by) the situation; i.e., $s,g\vdash \sigma\cdot pre$;
	\item $\sigma\cdot pre,\sigma \cdot post$ are grounded;
	\item $pre'$ is a set of grounded literals in $\sigma \cdot pre$ containing dynamic properties only;
	\item $post'=\sigma \cdot post$ and $con'=\sigma\cdot con$.
\end{itemize}
\end{definition}
Given an action $a$, we denote by $actor(a)$ the agent performing the action, and by $pre(a), con(a), post(a)$ the precondition, concurrent condition and postcondition.  

\begin{example}
In state $s_0$, the robot $r1$ moves from office $a$ to office $b$. This is formalised as follows:
\[\left\langle
\begin{array}{l}
move, \{robot(r1),\mathit{office(a)},\mathit{office(b)},in(r1,a),\\
corridor(a,b)\},\{\},\{\neg in(r1,a),in(r1,b)\}
\end{array}
\right\rangle\]
\end{example}

In a MAS, concurrent actions\footnote{Concurrent action means actions that occur at the same time and does not necessarily imply agent cooperation or coordination.} define state transitions. More formally, a concurrent action $A=\{ a_{1}, ... ,a_{n}\}$ is a set of individual actions. Given a set of actions $A=\{ a_{1}, ... ,a_{n}\}$, we define $pre(A)=\bigcup pre(a_{i})$, $post(A)=\bigcup post(a_{i})$ and $actor(A)=\bigcup actor(a_i)$.

Given a concurrent action $A=\{ a_{1}, ... ,a_{n}\}$ we say that the concurrent condition of an individual action $a_{i}$ of $A$ is \textit{satisfied} when for all positive schema in the concurrent condition exists an action $a_{j}$ ($i\neq j$) in $A$, such that $a_{j}$ is an instance of the schema; and for all negative schema none of the elements in $A$ is an instance of the schema. For the sake of simplicity, we assume that each agent performs one action at a time\footnote{This limitation can be relaxed by decomposing agents into groups of agents corresponding to agents' actuators \cite{boutilier2001partial}.}. 

\begin{definition}{(Consistency \cite{boutilier2001partial})}
Given a concurrent action $A=\{ a_{1},...,a_{n}\}$ it is \textit{consistent} if:
\begin{itemize}
	\item $pre(A)$ is consistent (i.e, $pre(A)\not\vdash \bot$);
	\item $post(A)$ is consistent (i.e, $post(A)\not\vdash \bot$);
	\item the concurrent condition of each action is satisfied;
	\item the concurrent action is complete (i.e., each agent performs one action in $A$).
\end{itemize}
\label{def:consistency}
\end{definition}
 
\begin{example}
The concurrent action $A=\{move(r1,a,b),\\move(r2,d,a),move(r3,e,a)\}$\footnote{For simplicity, we represent actions by their schemata.} is consistent since:
\begin{itemize}
	\item $pre(A)=\{in(r1,a),in(r2,d),in(r3,e)\}$ which is consistent;
	\item $post(A)=\{in(r1,b),in(r2,a),in(r3,a),\lnot in(r1,a),\allowbreak\lnot in(r2,d),\lnot in(r3,e)\}$ which is consistent;
	\item the concurrent conditions of both actions are satisfied;
	\item each robot performs one action. 
	\end{itemize}
\end{example}

A concurrent action $A=\{ a_{1}, ... ,a_{n}\}$ is applicable in a situation if $A$ is consistent and each individual action $a_{i}\in A$ is applicable in this situation. 

Given a consistent action, we define its effects as the postconditions of its individual actions and the preconditions not invalidated by the postconditions. More formally, given a concurrent action $A=\{ a_{1}, ... ,a_{n}\}$ its effects are a set of grounded literals as follows:
\[\mathit{eff}(A)=(\displaystyle\bigcup_{\substack{
            \forall pre\in pre(A):\\
            pre, post(A)\not\vdash \bot}}pre)\bigcup( \bigcup_{\forall post\in post(A)} post)\]

\subsection{Norm Definitions}
We consider \textit{norms} as formal statements that define patterns of behaviour by means of \textit{deontic modalities} (i.e., \textit{obligations} and \textit{prohibitions}). Specifically, our proposal is based on the notion of norm as a conditional rule of behaviour that defines under which circumstances a pattern of behaviour becomes relevant and must be fulfilled \cite{DBLP:journals/eaai/MeneguzziMOMLF12,vasconcelos2007resolving,y2006normative,dignum1999autonomous}.

\begin{definition}
A \textit{norm} is defined as a tuple $\langle deontic,$\\$condition,action\rangle$, where:
\begin{itemize}[noitemsep,topsep=0pt]
	\item $deontic\in\{\mathcal{O},\mathcal{P}\}$ is the deontic modality of the norm, determining if the norm is an obligation ($\mathcal{O}$) or prohibition ($\mathcal{P}$);
	\item $condition$ is a set of literals of $\mathcal{L}$ as well as equality and inequality constraints that represents the norm condition, i.e., it denotes the situations in which the norm is relevant.
	\item $action$ is a positive action schema that represents the action controlled by the norm.
\end{itemize}
\label{def:norm}
\end{definition}

\begin{example}
In our example, there is a norm that avoids collisions by forbidding any robot to move into an office when the office is occupied by another robot:
\[\langle \mathcal{P}, in(R1,L2), move(R2,L1,L2)\rangle\]
This norm states that when a robot $R1$ is located in office $O2$ other robots are forbidden to move from any office $L1$ to $L2$.
\end{example}

In line with related literature \cite{DBLP:conf/ijcai/AlechinaDL13,lorini2012logical,alechina2014norm}, we consider a \emph{closed legal system}, where everything is considered permitted by default, and obligation and prohibition norms define exceptions to this default permission rule. We also define that a norm is relevant to a specific situation if the norm condition is satisfied in the situation. Besides, we define that a norm condition is satisfied in a given situation when there is a substitution of the variables in the norm condition such that the constraints in the norm condition are satisfied and the positive (vs. negative) literals in the norm condition are true (vs. false) in the situation.

\begin{definition}
Given a specific situation denoted by a state $s$ and a set of static properties $g$, and a norm $\langle deontic,$\\$condition,action\rangle$; a \textit{norm instance} is a tuple $\langle deontic,action'\rangle$ such as:
\begin{itemize}[noitemsep,topsep=0pt]
	\item There is a substitution $\sigma$ such that the condition is satisfied in the situation; i.e., $s,g\vdash \sigma\cdot condition$;
	\item $action'=\sigma\cdot action$.
\end{itemize}
\end{definition}\vspace{-3pt}

\begin{example}
In state $s_{0}$ the norm that forbids robots to move into occupied offices is  instantiated as follows:
\begin{gather*}
\langle \mathcal{P},move(R2,L1,d)\rangle \text{ where $\sigma=\{L2/d\}$}\\
\langle \mathcal{P},move(R2,L1,a)\rangle \text{ where $\sigma=\{L2/a\}$}\\
\langle \mathcal{P},move(R2,L1,e)\rangle \text{ where $\sigma=\{L2/e\}$}
\end{gather*}
\end{example}

The semantics of instances (and norms in general) depends on their deontic modality. An obligation instance is fulfilled when the mandatory action is performed and violated otherwise, while a prohibition instance is violated when the forbidden action is performed and fulfilled otherwise. We classify detected violations (vs. fulfilments) into: \textit{identified} violations (vs. fulfilment), which refers to when the monitor knows the specific action that an agent executed and violates (vs. fulfils) an instance; and \textit{discovered} violations (vs. fulfilment), which refers to when the monitor knows that an agent violated (vs. fulfilment) some instance but does not know the forbidden (vs. mandatory) action executed by the agent. 

\section{NM Information Model}\label{sec:NM}

Let us assume a set of agents $Ag$ to be monitored, a set of norms $N$ that regulate the actions of agents, and a set $D$ of action descriptions that represent the actions that can be performed by agents. For the sake of simplicity, we assume that there is a single Norm Monitor (NM) that observes the actions performed by agents and monitors norm compliance\footnote{However, our model can be used by a team of monitors as well.}. We also assume that actions are deterministic and that the current state evolves due to action execution only\footnote{This assumption could be relaxed if NMs have capabilities for observing both state changes and actions.}. The goal of the NM is to analyse a partial sequence of action observations to detect norm violations. The enforcement of norms is out of the scope of this work and we assume that once the NM detects a norm violation (vs. fulfilment), it applies the corresponding sanction (vs. reward).
\subsection{State Representation}
As the NM only observes a subset of the actions performed by agents, it has partial information about the state of the world. The NM represents each partial state of the world, denoted by $p$, using an ``open world assumption'' as a set of grounded literals that are known in the state. Thus, a partial state contains positive (vs. negative) grounded literals representing dynamic properties known to be true (vs. false) in the state. The rest of dynamic properties are unknown.

To begin with, assume that the NM monitor has complete knowledge of the initial state (this will be relaxed later). Thus, at $t_{0}$ the NM knows which grounded atomic formulas are true or false in the initial state ($p_{0}\equiv s_{0}$). From that moment on, the NM monitors the actions performed by agents at each point in time. At time $t_{0}$ the NM carries out a monitoring activity and observes some of the actions performed by agents ($Act_{0}$). These actions have evolved $s_{0}$ into a new state $s_{1}$. As previously mentioned, the NM has limited capabilities for observing the actions performed by agents. Thus, it is possible that the NM observes a subset of the actions performed by agents. Specifically, if all actions have been observed ($|Act_{0}|=|Ag|$), then the resulting partial state $p_{1}$ can be constructed by considering the effects of actions in $Act_{0}$ on $p_{0}$ so $p_{1}\equiv s_{1}$. A different case arises when the NM observes a subset of the actions performed by the agents ($|Act_0|<|Ag|$). In this case, the agent cannot be sure about the effects of unobserved actions. Thus, the new partial state $p_{1}$ is constructed by assuming that the postconditions of the observed actions must hold on state $s_1$ (i.e., positive postconditions are positive literals in $p_1$ and negative postconditions are negative literals in $p_1$) and the rest of dynamic propositions are unknown. If the NM takes into account the next sequence of actions that it observes at time $t_{1}$ ($Act_1$), then the NM can also infer that the preconditions of these actions must hold on state $s_{1}$, and, as a consequence, new propositions can be taken for sure in the partial state $p_{1}$, retrospectively. Partial states in the general case are defined as:

\begin{definition}
Given a partial state description $p_t$ corresponding to time $t$, and two consecutive sequences of observed actions $Act_t$ and $Act_{t+1}$ executed by agents at times $t$ and $t+1$, respectively; the new partial state $p_{t+1}$ resulting from executing actions $Act_t$ in $p_t$ and actions $Act_{t+1}$ in $p_{t+1}$ is obtained as follows:
\[p_{t+1}=\left\{
\begin{array}{l l}
   post(Act_t) \bigcup pre(Act_{t+1}) & \text{if $|Act_t|<|Ag|$}\\
   p_t^{*} \bigcup \mathit{eff}(Act_t) \bigcup pre(Act_{t+1})  & \text{otherwise}\\
  \end{array} \right.
\]
where $p_t^{*}$ is the set of \textit{invariant} literals; i.e., literals of $p_t$ that have not been modified by the actions in $Act_t$ and it is defined as follows:
\[
p_t^{*}=\displaystyle\bigcup_{\substack{
            \forall l\in p_t:\\
            l,  \mathit{eff}(Act_t) \not\vdash \bot}} l
							\]
\label{def:staterec}
\end{definition}

\begin{example}
In our example, the NM knows which grounded atomic formulas are true or false in the initial state:
\begin{align*}
 p_0 = \{&in(r1,a),\lnot in(r1,b),\lnot in(r1,c),\lnot in(r1,d),\lnot in(r1,e),\\
&\lnot in(r1,f),in(r2,d),\lnot in(r2,a),\lnot in(r2,b),\lnot in(r2,c),\\
&\lnot in(r2,e),\lnot in(r2,f),in(r3,e),\lnot in(r3,a),\lnot in(r3,b),\\
&\lnot in(r3,c),\lnot in(r3,d),\lnot in(r3,f)\}
\end{align*}
The NM has some surveillance cameras to monitor the movement of robots in the building. Specifically, the corridors that are monitored are the ones between offices: $a$ and $b$; $b$ and $c$; and $b$ and $f$. These corridors are represented by black arrows in Figure \ref{fig:ExampleScenario}, whereas non-monitored  corridors are represented by grey arrows. In the initial state ($s_{0}$) depicted in Figure \ref{fig:initial}, the robots execute the actions $move(r1,a,b),move(r2,d,a)$ and $move(r3,e,a)$ resulting in a new state ($s_{1}$) depicted in Figure \ref{fig:first}. However, the NM only observes the action of robot $r1$, because this action takes place in a monitored corridor; i.e., $Act_{0}=\{move(r1,a,b)\}$. In the next state $s_{1}$, the robots execute actions $move(r1,b,c),move(r2,a,e)$ and $move(r3,a,b)$ resulting in a new state ($s_{2}$) depicted in Figure \ref{fig:second}. In this case the NM observes two actions; i.e., $Act_{1}=\{move(r1,b,c),move(r3,a,b)\}$. Considering these two sets of observed actions the NM is able to infer the dynamic propositions that are known in $s_{1}$ as follows:
\[p_{1}=\{in(r1,b),\lnot in(r1,a), in(r3,a)\}\]

If the NM uses the information about the states and the observed actions, then no violation of the norm is detected and no robot is sanctioned. However, $r2$ and $r3$ have violated the norm, since they have moved into an occupied office through non-monitored corridors.
\end{example}
\subsection{Action Reconstruction}
NMs use Definition \ref{def:staterec} to generate partial state descriptions based on the observed actions. Additionally, we propose that NMs reconstruct the actions that have not been observed. This reconstruction process entails: (i) searching for the actions that have been performed by unobserved agents; and (ii) using the actions found to increase the knowledge about the state of the world. The reconstruction process must be sound, e.g., it cannot indicate that a violation has occurred when it has not in fact occurred. In the following, we introduce full and approximate methods for reconstructing unobserved actions.

\subsubsection{Full Reconstruction}
Full reconstruction tries to find exhaustively the actions performed by all the agents that have not been observed. To this aim, the full reconstruction performs a search to identify all solutions to the reconstruction problem. 
\begin{definition}
Given a partial state description $p_t$ corresponding to time $t$ (named initial state), a set of observed actions $Act_{t}$ at time $t$, and an partial resulting state $p_{t+1}$ corresponding to time $t+1$ (named final state); we define \textit{search} as a function that computes sets of solutions $\mathcal{S}=\{S_{1},...,S_{k}\}$ such that each solution $S_{i}$ in $\mathcal{S}$ is a set of actions such that:
\begin{itemize}
	\item the concurrent action $S_{i}\cup Act_{t}$ is consistent;
	\item the initial state induced by the concurrent action $S_{i}\cup Act_{t}$ is consistent (i.e., $g,p_{t}, pre(S_{i}\cup Act_{t}), \triangledown\not\vdash \bot$);
	\item the final state induced by the concurrent action $S_{i}\cup Act_{t}$ is consistent (i.e., $g,p_{t+1}, post(S_{i}\cup Act_{t}), \triangledown\not\vdash \bot$).
\end{itemize}
\label{def:full}
\end{definition}
Thus, a solution is a set of actions performed by the agents that have not been observed\footnote{If all actions were observed, no reconstruction would be needed.} that are consistent with the states of the world before and after the execution of the actions. Given that the NM has a partial knowledge of the states, we do not require that the preconditions (vs. postconditions) of actions in a solution are met in the initial (vs. final) state, since it is possible that the preconditions (vs. postconditions) are true, but the NM is unaware of it. 

\begin{example}
Given the partial state description $p_0$, the set of observed actions $Act_{0}$, and the partial resulting state $p_{1}$, the search function looks for actions of agents $r2$ and $r3$ (since they are the agents that have not been observed). According to the initial position of $r2$, the NM can infer that $r2$ may have performed two different actions $move(r2,d,a)$ and $move(r2,d,e)$ ---these two actions are the only ones consistent with $p_0$. Similarly, the NM can infer that $r3$ may have performed three different actions $move(r3,e,a)$, $move(r3,e,d)$ and $move(r3,e,f)$ ---these three actions are the only ones consistent with $p_0$. However, the actions $move(r3,e,d)$ and $move(r3,e,f)$ are not consistent with the final state ---recall that these two actions have as postcondition the fact that $r3$ is in offices $d$ and $f$, respectively; that $p_{1}$ defines that $r3$ is in office $a$; and that $\triangledown$ defines as inconsistent states where any robot is at two different locations. As a result, the solution set for this problem is defined as:
\begin{align*}\mathcal{S}=\{\{move(r2,d,a),move(r3,e,a)\},\\\{move(r2,d,e),move(r3,e,a)\}\}\end{align*}
\end{example}

Once all solutions are found, the NM uses this information to extend the information about the actions performed by unobserved agents and the state of the world. To ensure that the reconstruction is sound, the NM calculates the intersection of actions in the solutions to select actions it is completely sure about (i.e., actions belonging to all solutions). Given a set of search solutions $\mathcal{S}=\{S_{1},...,S_{k}\}$ for some initial and final states, we define the \textit{reconstruction} action set  as follows:
\[R=\bigcap_{\forall S_{i}\in \mathcal{S}}S_{i}\]
If $R\neq\emptyset$, then the NM expands its knowledge about the actions performed by agents and it uses this information to increase the knowledge about the initial and final states. More formally, the set of actions observed in $t$ is updated as:
\[Act_t=Act_t\cup R\]
The initial state $p_t$ is updated as follows:
\[p_{t}=p_t \bigcup pre(R)
\] 
Finally, the final state is updated as follows:
\[p_{t+1}=\left\{
\begin{array}{l l}
    p_{t+1} \bigcup post(R)\bigcup p_t^\bullet  & \quad \text{if $|Act_t|<|Ag|$}\\
    p_{t+1} \bigcup p_t^{*} \bigcup  \mathit{eff}(Act_t)  & \quad \text{otherwise}\\
  \end{array} \right.
\]
where $p_t^{*}$ is defined as before and $p_t^\bullet$ is the set of extended invariant literals; i.e., literals in $p_t$ that have not been modified since there is not a solution $S_{i}\in \mathcal{S}$ such that the concurrent action $S_{i} \cup Act_t $ changes any of these literals:
\[
p_t^\bullet=\displaystyle\bigcup_{\substack{
            \forall l\in p_t:\\
            \nexists S_{i}\in \mathcal{S}: l,post(Act_t \cup S_{i}) \not\vdash \bot}} l
							\]

\begin{example}
The reconstruction set for the example is:
\[R=\{move(r3,e,a)\}\]
This action belongs to all solutions, so the NM can be absolutely sure about the performance of this action, even when the NM has not observed it. As a consequence, the NM extends its information as follows:
\[Act_{0}=\{move(r1,a,b),move(r3,e,a)\}\]
and $p_{0}$ remains unchanged and $p_1$ is updated as follows:
\begin{align*}
 p_1= \{&in(r1,b),\lnot in(r1,a),\lnot in(r1,c),\lnot in(r1,d),\lnot in(r1,e),\\
&\lnot in(r1,f),\lnot in(r2,b),\lnot in(r2,c),\lnot in(r2,f),in(r3,a),\\
&\lnot in(r3,b),\lnot in(r3,c),\lnot in(r3,d),\lnot in(r3,e),\lnot in(r3,f)\}
\end{align*}
\end{example}



The main disadvantage of full reconstruction is that, for many real-world problems, the number of candidate solutions that needs to be explored is prohibitively large, as shown later in Section \ref{sec:Alg}. In response to this problem, we provide a polynomial approximation below.

\subsubsection{Approximate Reconstruction}
Approximate reconstruction includes an approximate search that finds the actions performed by unobserved agents that are consistent with the states of the world before and after action execution. Specifically, approximate reconstruction identifies actions that do not necessarily include the specific actions performed by unobserved agents but that allow the NM to control norms. The main intuition beyond approximate reconstruction is as follows: imagine that at a given initial state an agent can perform just one action and that this action is forbidden (vs. mandatory). In this case, the NM \textit{identifies} that the agent has violated (vs. fulfilled) a norm. Besides that, if an agent can perform $n$ different actions and all these actions are forbidden (vs. mandatory), the NM does not need to know which action has been executed to conclude that a norm has been violated (vs. fulfilled)\footnote{Note that the propose of this paper is to monitor norms, not to determine whether agents are responsible for norm violations/fulfilments. Monitoring situations where agents can only execute forbidden/obligatory actions can help to detect norm-design problems. Additionally, the fact that an agent can only execute forbidden actions may be explained by the agent putting itself into these illegal situations (e.g., I am allowed to overtake but overtaking may put me in a situation where I can only exceed the speed limit). }. Hence, we say that a violation has been \textit{discovered} (instead of \textit{identified}). Given a set of prohibition instances $P$ and an action $a$, we define that the action $a$ is forbidden (denoted by $forbidden(P,a)$) when $\exists p\in P: \exists \sigma: \sigma\cdot action(p)=a$. Similarly, given a set of obligation instances $O$ and an action $a$, we define that the action $a$ is mandatory (denoted by $mandatory(O,a)$) when $\exists o\in O: \exists \sigma: \sigma\cdot action(o)=a$.

\begin{definition}
Given a partial state $p_t$, a set of observed actions $Act_{t}$ at time $t$, and a partial resulting state $p_{t+1}$; we define \textit{approximate search} as a function that calculates the set of all unobserved applicable actions $\widetilde{S}=\{a_i,...,a_n\}$ such that:
\begin{itemize}
	\item the preconditions of each action in $\widetilde{S}$ are consistent with the initial state (i.e., $\forall a_i\in \widetilde{S}: g,p_{t}, pre(a_i), \triangledown\not\vdash \bot$);
	\item the postconditions of each action in $\widetilde{S}$ are consistent with the final state (i.e., $\forall a_i\in \widetilde{S}: g,p_{t+1}, post(a_i), \triangledown\not\vdash \bot$);	
		\item actions in $\widetilde{S}$ are performed by unobserved agents (i.e., $actor(\widetilde{S})\cap actor(Act_t)=\emptyset$);
	\item all unobserved agents perform at least one action in $\widetilde{S}$ .
\end{itemize}
\end{definition}
\begin{example}
Given the partial state description $p_0$, the set of observed actions $Act_{0}$, and the partial resulting state $p_{1}$, the approximate search function looks for actions of agents $r2$ and $r3$ (since they are the agents that have not been observed). According to the initial position of $r2$, the NM can infer that $r2$ may have performed two different actions $move(r2,d,a)$ and $move(r2,d,e)$. Again, $r3$ may have performed action $move(r3,e,a)$. The approximate solution for this problem is defined as:
\[\widetilde{S}=\{move(r2,d,a),move(r2,d,e),move(r3,e,a)\}\]
\end{example}

As in full reconstruction, the NM uses approximate search solutions ($\widetilde{S}$) to expand its knowledge about the actions performed by unobserved agents and to increase the knowledge about the initial and final states. When an unobserved agent may have executed only one action, then the NM knows for sure that this action was executed. More formally, the \textit{reconstruction} action set is defined as follows:
\[ R=\displaystyle\bigcup_{\substack{\forall a\in \widetilde{S}:\not\exists a' \in \widetilde{S}:\\ a\neq a' \wedge actor(a)=actor(a')}} a\]
The set of actions observed in $t$ is updated as:
\[Act_t=Act_t\cup R\]
Then the initial state $p_t$ is updated as follows:
\[p_{t}=p_t \bigcup pre(R)
\] 
The final state is updated as follows:
\[p_{t+1}=\left\{
\begin{array}{l l}
    p_{t+1} \bigcup post(R)\bigcup p_t^\circ  & \quad \text{if $|Act_t|<|Ag|$}\\
    p_{t+1} \bigcup p_t^*\bigcup  \mathit{eff}(Act_t)& \quad \text{otherwise}\\
  \end{array} \right.
\]
where $p_t^*$ is defined as before and $p_t^\circ$ is the set of extended invariant literals in $p_t$; i.e., literals that have not been modified since there is not an observed action or an applicable action that changes them:
\[
p_t^\circ=\Bigg(\displaystyle\bigcup_{\substack{
            \forall l\in p_t:\\
            l,post(Act_t) \not\vdash \bot}} l \Bigg)
						\bigcap
						\Bigg(\displaystyle\bigcup_{\substack{
            \forall l\in p_t:\\
            l,post(\widetilde{S}) \not\vdash \bot}} l\Bigg)
							\]
Finally, the set of discovered violations and fulfilments is a set of actions defined as follows:
\[ D=\{a_{1},...,a_{j}\}\]
where for each action $a_i$ in $D$: $a_i$ is in $\widetilde{\mathcal{S}}$ and the agent that performs $a_i$ (i.e., $actor(a_i)$ ) is:
\begin{itemize}	
	\item able to execute more than one action (i.e., $\exists a_j\in\widetilde{S}:\allowbreak a_i\neq a_j \wedge actor(a_i)=actor(a_j)$);
	\item only able to execute forbidden (vs. mandatory) actions and $a_i$ is one of these forbidden (vs. mandatory) actions;
\end{itemize}
When an agent is only able to perform forbidden (vs. mandatory) actions, an action among these can be selected according to various criteria. For example, in a normative system where the presumption of innocence principle holds, the NM should assume that the agent has violated (vs. fulfilled) the least (vs. most) important norm and the action that violates (vs. fulfils) this norm is selected. Note discovering violations is very useful in many practical applications, in which it would allow the NM to ban offender agents (e.g., Intrusion Detection/Prevention Systems \cite{bass2000intrusion}), to stop the execution of any offender agent (e.g., Business
Process Compliance monitoring \cite{sadiq2007modeling}), or to put offender agents under close surveillance (e.g., Model-Based Diagnosis Systems \cite{micalizio2004line}), even when the specific action performed ins not known.


\begin{example}
In case of the approximate reconstruction, $r3$ is only able to perform one action, which entails that the NM can be absolute sure about the performance of this action and the reconstruction set is defined as:
\[R=\{move(r3,e,a)\}\]
As a consequence, the NM extends its information as follows:
\[Act_{0}=\{move(r1,a,b),move(r3,e,a)\}\]
$p_0$ remains unchanged and $p_1$ is updated as follows:
\begin{align*}
 p_1= \{&in(r1,b),\lnot in(r1,a),\lnot in(r1,c),\lnot in(r1,d),\lnot in(r1,e),\\
&\lnot in(r1,f),\lnot in(r2,b),\lnot in(r2,c),\lnot in(r2,f),in(r3,a),\\
&\lnot in(r3,b),\lnot in(r3,c),\lnot in(r3,d),\lnot in(r3,e),\lnot in(r3,f)\}
\end{align*}
In this situation, $r2$ is only able to execute forbidden actions ---recall that the instances $\langle \mathcal{P},move(R2,L1,a)\rangle$ and $\langle \mathcal{P},move(R2,L1,e)\rangle$ forbid any robot to move into offices $a$ and $e$ and that $r2$ may have been executed actions $move(r2,d,a)$ and $move(r2,d,e)$. Thus, the set of discovered violations and fulfilments is defined as follows:
 \[D=\{move(r2,d,e)\}\]
note that the discovered violation does not correspond to the action executed by $r2$, however, it allows the NM to determine that $r2$ must have violated an instance.
\end{example}
\subsection{Norm Monitoring}
Once all the information about the actions performed by the agents and the partial states has been reconstructed, the NM checks the actions of agents to determine which instances have been violated or fulfilled. Recall that norms in our model are defined as conditional rules that state which actions are obligatory or forbidden. Given that the NM has partial knowledge about the state of the world, the NM should control norms only when it is completely sure that the norms are relevant to ensure that the norm monitoring process is sound. In particular, we define that a norm is relevant to a partial situation when the norm condition is satisfied by the partial situation ---i.e., a norm $\langle deontic, condition,  action\rangle$ is relevant to a partial situation represented by a partial state $p$, the static properties $g$ and the domain knowledge $\triangledown$ if $\exists \sigma$ such that $p,g,\triangledown \vdash \sigma \cdot condition$. 

\begin{example}
In state $p_{0}$ the norm that forbids robots to move into occupied offices is  instantiated three times as follows:
\begin{gather*}
\langle \mathcal{P},move(R2,L1,d)\rangle \text{ where $\sigma=\{L2/d\}$}\\
\langle \mathcal{P},move(R2,L1,a)\rangle \text{ where $\sigma=\{L2/a\}$}\\
\langle \mathcal{P},move(R2,L1,a)\rangle \text{ where $\sigma=\{L2/e\}$}
\end{gather*}
\end{example}

Once the NM has determined which norm instances hold in a given situation, it has to check the actions of agents to determine which instances have been violated and which ones have been fulfilled.

\textit{Obligation Instance}. In presence of partial knowledge about the actions performed by agents, the NM can only determine that an obligation instance has been fulfilled. If the NM knows all the actions performed by agents, then it can determine whether an obligation has been fulfilled or violated.

	\begin{definition}
	Given an obligation instance $\langle \mathcal{O},action'\rangle$ and a set of observed actions $Act$, then the obligation is defined as:
	\[\left\{
  \begin{array}{l l}
    fulfilled & \quad \text{iff $\exists \sigma: \sigma\cdot action'\in Act$}\\
    violated & \quad \text{iff $(\not\exists \sigma: \sigma\cdot action'\in Act) \wedge|Act|=|Ag|$}\\
    unknown & \quad \text{otherwise}
  \end{array} \right.
	\]
	\label{def:obligation}
	\end{definition}

\textit{Prohibition Instance}. In presence of partial knowledge about the actions performed by agents, the NM can only determine that a prohibition instance has been violated. If the NM knows all the actions performed by agents then it can determine whether a prohibition has been fulfilled or violated.

	\begin{definition}
	Given a prohibition instance $\langle \mathcal{P},action'\rangle$ and a set of observed actions $Act$, then the prohibition is defined as:
	\[\left\{
  \begin{array}{l l}
    violated & \quad \text{iff $\exists \sigma: \sigma\cdot action'\in Act$}\\
    fulfilled & \quad \text{iff $(\not\exists \sigma: \sigma\cdot action'\in Act) \wedge|Act|=|Ag|$}\\
    unknown & \quad \text{otherwise}
  \end{array} \right.
	\]
	
	\label{def:prohibition}
	\end{definition}

Finally, the set of discovered violations and fulfilments is used to identify those agents that have violated or fulfilled an instance. 

\begin{example}
Taking into account the set of actions $Act_{0}$, the NM can \emph{identify} that robot $r3$ has violated the instance $\langle \mathcal{P},move(R2,L1,a)\rangle$, even though this forbidden action has not been observed by the NM. Specifically, there is $\sigma=\{R2/r3, L1/e\}$ such that $\sigma(move(R2,L1,a)) \in Act_{0}$. Besides that, the approximate reconstruction \emph{discovers} that robot $r2$ has violated a prohibition instance though it doe snot know the exact action performed---recall that $D=\{move(r2,d,e)\}$. Had the NM not performed the proposed reconstruction processes, none of these violations would have been detected. 
\end{example}

\section{NM Algorithms}\label{sec:Alg}
Algorithm \ref{alg:monitor} contains the NM pseudocode. In each step, the NM observes the actions of agents and uses this information to update the current and the previous partial states (lines 4-9). If all the actions have not been observed in the previous state, then the NM executes the $reconstruction$ function to reconstruct unobserved actions (lines 11-14). Then, the $checkNorms$ function is executed to determine which norms have been violated and fulfilled in the previous state (line 15) according to Definitions \ref{def:obligation} and \ref{def:prohibition}.

Note that the NM code can be executed while actions are performed without delaying agents. 
Regarding the temporal cost of the algorithm executed by NMs, it is determined by the cost of the $reconstruction$ function, the implementations of which (full and approximate) are  discussed below.

\begin{algorithm}
\begin{small}
\begin{algorithmic}[1]
\Require $Ag, N, D, \triangledown, g$
\State $p_{0}=\emptyset$ \Comment{$p_{0}$ is an empty conjunction of literals}
\State $t\gets 0$
\While {$true$}
\State $Act_{t}\gets observeActions()$
\If{$|Act_{t}|<|Ag|$}
\State $p_{t+1}\gets post(Act_{t})$
\Else
\State $p_{t+1}\gets p^*_{t} \bigwedge  \mathit{eff}(Act_{t})$
\EndIf
\State $p_{t}\gets p_{t} \bigwedge pre(Act_{t})$
\If{$t>0$}
\If{$|Act_{t-1}|<|Ag|$}
\State $TA\gets Ag\setminus actors(Act_{t-1})$\Comment{Target Agents}
\State $D\gets \emptyset$\Comment{Discovered violations and fuflilments}
\State $reconstruction(p_{t-1},p_{t},Act_{t-1},TA,D)$
\EndIf
\State $checkNorms(p_{t-1},Act_{t-1})$
\EndIf
\State $t\gets t+1$
\EndWhile
\end{algorithmic}
\caption{NM Algorithm}\label{alg:monitor}
\end{small}
\end{algorithm}

\smallskip
\textit{Full Reconstruction (Algorithm \ref{alg:solve}).} This pseudocode corresponds to the full reconstruction function. This function calls the function $search$ to search the actions of target agents (line 2). Then, for all the solutions found, the NM checks if they are consistent according to Definition \ref{def:full} (lines 4-6). Finally, consistent solutions are used to extend the set of observed actions and the knowledge about the initial and final states (lines 7-14). The temporal cost of this algorithm is given by the cost of the $search$ function discussed below.

\begin{algorithm}
\begin{small}
\begin{algorithmic}[1]
\Function{FullReconstruction}{$i,f,Act,TA,D$}
\State {$\mathcal{S'}\gets search(i,f,Act,TA)$} \Comment{Candidate Solutions}
\State $\mathcal{S}\gets\emptyset$ \Comment{Consistent Solutions}
\ForAll{$S_{j}\in \mathcal{S'}$}
\If{$checkSolutionConsistency(Act,S_{j},i,f)$}
\State $\mathcal{S}\gets\mathcal{S}\cup S_{j}$
\EndIf
\EndFor
\State $R\gets \bigcap_{\forall S_{i}\in \mathcal{S}} S_{i}$
\If{$R\neq\emptyset$}
\State $Act\gets Act\cup R$
\State $i\gets i\bigwedge pre(R)$
\If{$|Act|<|Ag|$}
\State $f\gets f\bigwedge post(R)\bigwedge i^\bullet$
\Else
\State $f\gets i^*\bigwedge f\bigwedge  \mathit{eff}(Act) $
\EndIf
\EndIf
\EndFunction
\end{algorithmic}
\caption{Full Reconstruction Function}\label{alg:solve}
\end{small}
\end{algorithm}

Algorithm \ref{alg:search} contains the pseudocode of the recursive $search$ function that computes all the sequences of consistent actions that may have been executed by the agents that have not been observed. It starts by checking that there is at least one target agent (line 2). If so, it identifies all actions that might have been executed by one target agent (lines 3-4). An action might have been executed if it is consistent according to the static properties, the domain knowledge, and the initial and final states. For each consistent action, it reconstructs the actions of the remaining agents recursively (lines 5-13). In the worst case, the temporal cost of this function is $O(|Ag|^{|D|\times I_{D}})$, where $Ag$ is the set of agents, $D$ is the set of action descriptions and $I_{D}$ is the maximum number of instantiations per action. This situation arises when no action is observed and all actions are applicable for all agents. 

\begin{algorithm}
\begin{small}
\begin{algorithmic}[1]
\Function {search}{$i,f,Act,TA$}
\If {$TA\neq \emptyset$}
\ForAll {$d\in D$}
\Comment Identify consistent actions
\If {$\exists \sigma: checkActionConsitency(\sigma \cdot d,i,f) \wedge actor(\sigma\cdot d)\in TA$}
\State $\alpha=actor(\sigma\cdot d)$
\State $i'\gets i \bigwedge pre(\sigma \cdot d)$
\State $f'\gets f \bigwedge post(\sigma\cdot d)$
\State $Act'\gets Act \cup \sigma\cdot d$
\State $TA'\gets TA \setminus \{\alpha\}$
\State $\mathcal{S}\gets search(i',f',Act',TA')$
\ForAll{$S_i\in \mathcal{S}$}
\State $S_i\gets S_i\cup\sigma\cdot d$
\EndFor
\State \Return $\mathcal{S}$
\EndIf
\EndFor
\Else
\State \Return $\emptyset$
\EndIf
\EndFunction
\end{algorithmic}
\caption{Search Function}\label{alg:search}
\end{small}
\end{algorithm}

\smallskip
\textit{Approximate Reconstruction Function (Algorithm \ref{alg:areconstruct})}. This function calls the function $ApproximateSearch$ to search the applicable actions per each target agent (line 2). Then, the list of applicable actions per each agent is checked (lines 3-12). Specifically, if an agent may have executed one action only, then the NM knows that this action was executed and it updates the reconstructed action set (lines 3-5). Then, the set of observed actions and the knowledge about the initial and final states is updated (lines 6-12). Finally, discovered violations and fulfilments are calculated (lines 14-19). The temporal cost of this algorithm is given by the cost of the $ApproximateSearch$ function discussed below.

Algorithm \ref{alg:asearch} contains the pseudocode of the $ApproximateSearch$ function. It starts by initialising the list of applicable actions per agent (lines 3-4). Then it calculates the set of instances that are relevant to the initial state (line 7). The function calculates per each target agent the list of applicable actions that it may have executed (lines 9-11). Then, the list of applicable actions per each agent is checked (lines 12-18). Specifically, if an agent may have executed one action only, then the NM knows that this action was executed and it updates the list of applicable actions, the initial and final states, and retracts the agent from the target agents (lines 14-17). This process is repeated until there are no more target agents or the initial and final states remain unchanged. Then the set of instances that are relevant to the initial state is calculated (line 13). Finally, the list of applicable actions per agent is updated with actions of remaining target agents (lines 19-20). The temporal cost of this function is $O(|Ag|^2\times|D|\times I_D)$.

\begin{algorithm}
\begin{small}
\begin{algorithmic}[1]
\Function{ApproximateReconstruction}{$i,f,Act,TA,D$}
\State{$\widetilde{S}\gets approximateSearch(i,f,Act,TA)$}
\ForAll{$\alpha\in TA$}
\If{$|\widetilde{S}_{\alpha}|=1$}
\State $R\gets R \cup S_{\alpha}$
\EndIf
\EndFor
\If{$R\neq\emptyset$}
\State $Act\gets Act\cup R$
\State $i\gets i\bigwedge pre(R)$
\If{$|Act|<|Ag|$}
\State $f\gets f\bigwedge post(R)\bigwedge i^\circ$
\Else
\State $f\gets i^*\bigwedge f\bigwedge  \mathit{eff}(Act) $
\EndIf
\EndIf
\State $O,P\gets calculateInstances(i)$
\ForAll{$\alpha\in TA$}
\If{$|\widetilde{S}_{\alpha}|>1$}
\If{$\not\exists a\in \widetilde{S}_{\alpha}: \lnot forbidden(P,a)$}
\State $D\gets D\cup a$\Comment{$a$ is an action from $\widetilde{S}_{\alpha}$}
\ElsIf{$\not\exists a\in \widetilde{S}_{\alpha}: \lnot mandatory(O,a)$}
\State $D\gets D\cup a$ \Comment{$a$ is an action from $\widetilde{S}_{\alpha}$}
\EndIf
\EndIf
\EndFor
\EndFunction
\end{algorithmic}
\caption{Approximate Reconstruction Function}\label{alg:areconstruct}
\end{small}
\end{algorithm}

\begin{algorithm}
\begin{small}
\begin{algorithmic}[1]
\Function{ApproximateSearch}{$i,f,Act,TA$}
\State{$continue\gets true$}
\ForAll{$\alpha\in TA$}
\State{$\widetilde{S}_{\alpha}\gets \emptyset$}\Comment{List of approximate actions per agent}
\EndFor
\While{$continue\wedge TA\neq\emptyset$}
\State{$continue\gets false$}
\ForAll{$\alpha\in TA$}
\State $L_{\alpha}\gets \emptyset$
\EndFor
\ForAll {$d\in D$}
\If {$\exists \sigma: checkActionConsitency(\sigma\cdot d,i,f) \wedge actor(\sigma\cdot d)\in TA$}
\State $L_{actor(\sigma\cdot d)}\gets L_{actor(\sigma\cdot d)}\cup \sigma\cdot d$
\EndIf
\EndFor
\ForAll{$\alpha\in TA$}
\If{$|L_{\alpha}|=1$}
\State $\widetilde{S}_{\alpha}\gets L_{\alpha}$
\State $i\gets i\bigwedge pre(L_{\alpha})$
\State $f\gets f\bigwedge post(L_{\alpha})$
\State $TA\gets TA\setminus \{\alpha\}$
\State $continue\gets true$
\EndIf
\EndFor
\EndWhile
\ForAll{$\alpha\in TA$}
\State {$\widetilde{S}_{\alpha}\gets L_{\alpha}$}
\EndFor
\State \Return $\widetilde{S}$
\EndFunction
\end{algorithmic}
\caption{Approximate Search Function}\label{alg:asearch}
\end{small}
\end{algorithm}

\section{Evaluation}\label{sec:eva}
This section compares the performance of a NM with full reconstruction, a NM with approximate reconstruction and a traditional norm monitor ---which is the method used in the majority of
previous proposals \cite{cardoso2007institutional,minsky2000law,gaertner2007distributed,modgil2009framework,criado2013manea}--- that only considers the observed actions to detect violations; with respect to their capabilities to monitor norm compliance. We have evaluated our proposal in a case study, which allows us to contextualise the results and to give a meaningful interpretation to them; and in a series of random experiments, which allow us to evaluate our proposal under a wide range of different situations and parameter values.

\subsection{Case Study}
We implemented in Java a simulator of the paper example in which robots attend requests in offices connected through corridors. Compliance with the collision avoidance norm is controlled by a monitor that observes surveillance cameras. In each simulation, we generate corridors and cameras randomly. In each step of the simulation, each robot chooses randomly one applicable action to be executed. The simulation is executed $100$ steps and repeated $100$ times to support the findings. We conducted experiments in which the number of offices $O$ took a random value within the $\llbracket 3,500\rrbracket$ interval and the number of robots $R$ took a random value within the $\llbracket 2,250\rrbracket$ interval. Besides that, to be able to compare with the full NM, we also considered small scenarios only, in which the number of offices $O$ takes a random value within $\llbracket 3,10 \rrbracket$ and the number of robots $R$ takes a random value within $\llbracket 2,5\rrbracket$, as the full reconstruction has an exponential cost and it is intractable for most of the cases with the default intervals.

\subsubsection{Action Observability}To analyse the performance and scalability of monitors with respect to their capabilities to observe actions, we defined the number of corridors $C$ as a random value within the $\llbracket O,O\times(O-1)\rrbracket$ interval and varied the ratio of cameras to corridors (action observability). Table \ref{tab:obs} shows the percentage of violations detected per each type of monitor. The higher the ratio of cameras, the more actions are observed and the better the performance of all monitors. Moreover, the approximate NM offers on average a $39\%$ performance improvement over a traditional monitor (i.e., it identifies 16\% more violations plus a further 24\% of discovered violations). That is, an approximate NM outperforms a traditional monitor with the same capabilities to observe actions. When compared to full NM in small scenarios ($O\in\llbracket 3,10\rrbracket$ and $R\in\llbracket 2,5\rrbracket$), approximate NM performs similarly. This is explained by the fact that there is a single norm in this scenario, actions have no concurrency conditions, and the preconditions and postconditions of actions are disjoint. In this circumstances, the approximate reconstruction process reconstructs actions similarly to the full reconstruction\footnote{Note that full reconstruction does not guarantee completeness.}. 



\begin{table}[h]
\center{
\scriptsize{\hspace{-3pt}
\begin{tabular}{ |c|c|c| }
\hline
Cameras&Traditional&Approximate NM\\
Ratio&Monitor&Identify+Discover\\
\hline
0\%&	0\%&		0+0\%\\ \hline
20\%&	11\%&		14+9\%\\ \hline
40\%&	31\%&		40+12\%\\ \hline
60\%&	55\%&		71+10\%\\ \hline
80\%&	78\%&		91+4\%\\ \hline
100\%&	100\%&		100+0\%\\ \hline
\multicolumn{3}{c}{$O\in\llbracket 3,500\rrbracket$ and $R\in\llbracket 2,250\rrbracket$}\\
\end{tabular}
\begin{tabular}{ |c|c|c|c| }
\hline
Cameras&Traditional&Full&Approximate NM\\
Ratio&Monitor&NM&Identify+Discover\\
\hline
0\%&	0\%&	0\%&	0+0\%\\ \hline
20\%&	16\%&	32\%&	32+6\%\\ \hline
40\%&	32\%&	68\%&	67+5\%\\ \hline
60\%&	56\%&	88\%&	88+3\%\\ \hline
80\%&	76\%&	99\%&	99+0\%\\ \hline
100\%&	100\%&	100\%&	100+0\%\\ \hline

\multicolumn{4}{c}{$O\in\llbracket 3,10\rrbracket$ and $R\in\llbracket 2,5\rrbracket$}\\
\end{tabular}}}

\caption{Action Observability Experiment}
\label{tab:obs}

\end{table}

\subsubsection{Action Instantiations}To analyse the performance and scalability of monitors with respect to agent capabilities to execute actions (i.e., the number of instantiations per action), we varied the ratio of corridors\footnote{Recall that $C$ takes values within the $\llbracket O,O\times(O-1)\rrbracket$ interval.} (e.g., a ratio of 0\% means $C=O$) and defined the number of cameras as a random value within the $\llbracket 0,C\rrbracket$ interval. Table \ref{tab:inst} shows the results of this experiment. The approximate NM offers on average a $43\%$ performance improvement over a traditional monitor (i.e., it identifies 29\% more violations plus a further 14\% of discovered violations). That is, given the same number of possible instantiations per action, an approximate NM outperforms a traditional monitor. Besides, we can see that, as in the previous experiment, the approximate NM performs similarly to the full NM. In particular, when the ratio of corridors is higher than 0\%, agents are capable of executing different actions and the reconstruction process becomes more complex, which decreases the performance of full and approximate NMs. However, full and approximate NMs noticeably outperform the traditional monitor regardless of the ratio of corridors.


\begin{table}[h]
\center{
\scriptsize{\hspace{-3pt}\begin{tabular}{ |c|c|c| }
\hline
Corridors&Traditional&Approximate NM\\
Ratio&Monitor&Identify+Discover\\
\hline
0\%&	51\%&		98+0\%\\ \hline
20\%&	48\%&		55+6\%\\ \hline
40\%&	48\%&		56+7\%\\ \hline
60\%&	41\%&		47+6\%\\ \hline
80\%&	49\%&		56+13\%\\ \hline
100\%&	42\%&		49+7\%\\ \hline
\multicolumn{3}{c}{$O\in\llbracket 3,500\rrbracket$ and $R\in\llbracket 2,250\rrbracket$}\\
\end{tabular}
\begin{tabular}{ |c|c|c|c| }
\hline
Corridors&Traditional&Full&Approximate NM\\
Ratio&Monitor&NM&Identify+Discover\\
\hline
0\%&	52\%&	99\%&	99+0\%\\ \hline
20\%&	59\%&	80\%&	79+3\%\\ \hline
40\%&	55\%&	74\%&	74+3\%\\ \hline
60\%&	51\%&	70\%&	69+4\%\\ \hline
80\%&	55\%&	68\%&	68+5\%\\ \hline
100\%&	57\%&	70\%&	69+4\%\\ \hline
\multicolumn{4}{c}{$O\in\llbracket 3,10\rrbracket$ and $R\in\llbracket 2.,5\rrbracket$}\\
\end{tabular}
}}

\caption{Action Instantiations Experiment}

\label{tab:inst}
\end{table}
\subsection{Random Experiments}
We implemented a simulator in Java in which there is a set of agents that perform actions in a monitored environment as defined below. In particular, our simulator does not model a specific scenario; rather it creates a different scenario in each simulation (i.e., generating randomly agent capabilities, the environment properties, actions and norms). 
As in the previous experiments, we have considered big and small scenarios. In particular, the number of agents $G$ in small scenarios took a random value within the $\llbracket 1,5 \rrbracket$ interval, whereas in big scenarios  $G$ took a random value within the $\llbracket 1, 500\rrbracket$ interval. The number of actions $A$ took a random value within the $\llbracket 1,50\rrbracket$ interval. Again, the simulation is executed $100$ steps and repeated $1000$ times to ensure that the values of the simulation parameters range over possible values\footnote{Note that in the random experiments there are more simulation parameters than in the case-study simulator and a higher number of repetitions is required to support the findings.}. 

\textit{Agent Definition. }We modelled different types of agents with different capabilities to perform actions. In particular, the set of actions available to each agent depends on the function/s assumed by each agent in a particular simulation. To model these capabilities, a set of roles is created at the beginning of each simulation. Specifically, the number of roles created took a random value within the $\llbracket 1, A\rrbracket$ interval. For each role a subset of the actions are randomly selected as the role capabilities; i.e., all agents enacting this role are able to perform these actions\footnote{This condition has been formulated in action preconditions as explained below.}. To avoid that all roles have similar capabilities, which would lead to simulations populated by homogeneous agents, the number of actions selected as role capabilities took a random value within the $\llbracket 1,\lceil0.1*A\rceil\rrbracket$ interval (i.e., at maximum each role is capable of performing a 10\% of the actions). At the beginning of each simulation, each agent is defined as enacting a random subset of the roles. In each step of the simulation, each agent selects randomly one action among the available actions that it can execute in the current state.

\textit{Environment Definition. }In the simulator, the environment is described in terms of different situations or states of affairs that can be true or false. Each one of these states of affairs is represented by a grounded proposition. Thus, the state of the environment is defined in terms of a set of propositions. For simplicity, we assumed that these propositions are independent (i.e, propositions are not logically related). In our simulations, the number of propositions $P$ took a random value within the $\llbracket A,2*A \rrbracket$ interval (i.e., there is at least one proposition per each action\footnote{Note that an action can change the truth value of several propositions.}). Besides that, there is a set of grounded atomic formulas describing the roles played by agents and the actions that can be performed by each role. The relationship between agents and roles is formally represented by a binary predicate ($play$). Specifically, the expression $play(g,r)$ describes the fact that the agent identified by $g$ enacts the role identified by $r$. Similarly, relationship between roles and actions is formally represented by a binary predicate ($capable$). Specifically, the expression $capable(a,r)$ describes the fact that agents enacting role $r$ are capable of performing the action identified by $a$. For simplicity, we assume that the roles enacted by the agents and the role capabilities are static properties of the environment.   

\textit{Action Definition.} Actions allow agents to change the state of the environment. At the beginning of each simulation, a set of actions is randomly generated. For each action $\langle name,pre,con,post\rangle$ the elements are defined as follows: $name$ is initialised with a sequential identifier $a$; $pre$ is defined as $\{play(A,R),capable(R,a),p_1,...,p_n\}$ where the elements $p_1,...,p_n$ are randomly selected from the proposition set; $con$ is defined as $\{a_1(A_1,R_1),...,a_m(A_m,R_m)\}$, where each $a_i$ is an action randomly selected from the action set such that $a_i\neq a$ and $A_i,R_i$ are free variables representing the agent performing the action and the role capable of performing this action, respectively; and $post$ is defined as $\{p_1,...,p_k\}$ where each $p_i$ is a proposition randomly selected from the proposition set. To avoid that actions have too many constraints, which would be unrealistic and make actions to be only executed on few situations, the number of propositions in $pre$ and $post$ takes a random value within the $\llbracket 1,\lceil 0.1*P\rceil\rrbracket$ interval. Similalry, the number of actions in $con$ takes a random value within the $\llbracket 0,\lceil 0.1*A\rceil\rrbracket$ interval. 
 
Besides these actions, a \textsf{NOP} action, which has no effect on the environment, was created. To maximise the number of actions executed in the simulations, which may entail more violations and fulfilments, we defined that the \textsf{NOP} action can only be executed by agents when none of their available actions can be executed. However, similar results would have been obtained if this condition was relaxed\footnote{Note that the capabilities of monitors to detect violations and fulfilments do not depend on the fact that agents are allowed to perform the \textsf{NOP} action in any situation.}. Our simulator models scenarios where the \textsf{NOP} action can always be observed. This is the case in many real domains such as Intrusion Detection Systems or Autonomous Systems, where it it not always possible to analyse the data (e.g., the packages) sent by agents (e.g., hosts) to infer the actions performed, but it is always possible to know which agents have performed an action (i.e., which agents have sent packages). 
 
\textit{Norm Definition.} Agents' actions are regulated by a set of norms. At the beginning of each simulation, a set of norms is randomly created. In particular the number of norms took a random value within the $\llbracket 1, A\rrbracket$ (i.e., there is at maximum one norm per each action). For each action $\langle deontic,condition,action\rangle$ the elements are defined as follows: $deontic$ is randomly initialised with a deontic operator; $condition$ is defined as $\{p_1,...,p_k\}$ where each $p_i$ is a proposition randomly selected from the proposition set; and $action$ is randomly initialised with an action. To allow norms to be instantiated, the number of propositions in $condition$ takes a random value within the $\llbracket 0,\lceil P*0.1\rceil\rrbracket$ interval.

\subsubsection{Action Observability}To analyse the performance and scalability of monitors with respect to their capabilities to observe actions, we varied the observation probability. Tables \ref{tab:obsF} and \ref{tab:obsV} show the percentage of detected fulfilments and violations, respectively. Again, the approximate NM offers a significant performance improvement over a traditional monitor; i.e., the approximate NM offers on average a 74\% performance improvement over a traditional monitor. When compared to full NM in small scenarios ($A\in \llbracket 1,50\rrbracket$ and $G\in\llbracket 1,5\rrbracket$), the full NM offers on average a $21\%$ performance improvement over an approximate NM. This is explained by the fact that this experiment is more complex than the case study; i.e., there are several norms (both prohibition and obligation norms), actions have concurrent conditions and actions may have conflicting preconditions and postconditions (i.e., conditions that are defined over the same propositions). Note that the traditional monitor detects violations and fulfilments even when the observation probability is 0\%. These detections correspond to situations in which none of the agents can execute any action (i.e., all agents execute the \textsf{NOP} action) which leads to the fulfilment of prohibition instances and the violation of obligation instances. This phenomenon is more frequent in case of small scenarios since the lower the number of agents, the higher the probability that all agents cannot execute any action.

\begin{table}[h]
\center{
\scriptsize{\hspace{-3pt}
\begin{tabular}{ |c|c|c| }
\hline
Observ.&Traditional&Approximate NM\\
Prob.&Monitor&Identify+Discover\\
\hline
0\%&	2\%&		35+7\%\\ \hline
20\%&	18\%&		47+6\%\\ \hline
40\%&	35\%&		62+4\%\\ \hline
60\%&	50\%&		72+3\%\\ \hline
80\%&	66\%&		80+2\%\\ \hline
100\%&	100\%&		100+0\%\\ \hline
\multicolumn{3}{c}{$A\in\llbracket 1,50\rrbracket$ and $G\in\llbracket 1,500\rrbracket$}\\
\end{tabular}
\begin{tabular}{ |c|c|c|c| }
\hline
Observ.&Traditional&Full&Approximate NM\\
Prob.&Monitor&NM&Identify+Discover\\
\hline
0\%&	20\%&	46\%&	34+1\%\\ \hline
20\%&	23\%&	51\%&	39+1\%\\ \hline
40\%&	31\%&	57\%&	45+1\%\\ \hline
60\%&	43\%&	69\%&	56+0\%\\ \hline
80\%&	58\%&	79\%&	70+0\%\\ \hline
100\%&	100\%&	100\%&	100+0\%\\ \hline
\multicolumn{4}{c}{$A\in\llbracket 1,50\rrbracket$ and $G\in\llbracket 1,5\rrbracket$}\\
\end{tabular}}}

\caption{Fulfilments Detected in the Action Observability Experiment}
\label{tab:obsF}
\end{table}

\begin{table}[h]
\center{
\scriptsize{\hspace{-3pt}
\begin{tabular}{ |c|c|c| }
\hline
Observ&Traditional&Approximate NM\\
Prob.&Monitor&Identify+Discover\\
\hline
0\%&	2\%&		36+8\%\\ \hline
20\%&	18\%&		50+6\%\\ \hline
40\%&	35\%&		62+6\%\\ \hline
60\%&	49\%&		70+2\%\\ \hline
80\%&	65\%&		78+2\%\\ \hline
100\%&	100\%&		100+0\%\\ \hline
\multicolumn{3}{c}{$A\in\llbracket 1,50\rrbracket$ and $G\in \llbracket 1,500\rrbracket$}\\
\end{tabular}
\begin{tabular}{ |c|c|c|c| }
\hline
Observ&Traditional&Full&Approximate NM\\
Prob.&Monitor&NM&Identify+Discover\\
\hline
0\%&	20\%&	44\%&	33+1\%\\ \hline
20\%&	23\%&	50\%&	37+0\%\\ \hline
40\%&	32\%&	56\%&	45+0\%\\ \hline
60\%&	42\%&	68\%&	55+0\%\\ \hline
80\%&	56\%&	78\%&	69+0\%\\ \hline
100\%&	100\%&	100\%&	100+0\%\\ \hline
\multicolumn{4}{c}{$A\in\llbracket 1,50\rrbracket$ and $G\in\llbracket 1,5\rrbracket$}\\
\end{tabular}}}

\caption{Violations Detected in the Action Observability Experiment}
\label{tab:obsV}
\end{table}
\subsubsection{Action Possibilities}To analyse the performance and scalability of monitors with respect to agent capabilities to execute actions (i.e., the number of available actions), we defined the observation probability as a random value within the $[0,100\%]$ interval and we varied the number of actions. Tables \ref{tab:instF} and \ref{tab:instV} show the percentage of detected fulfilments and violations, respectively. In this experiment, the more actions, the more complex the reconstruction problem is. As a consequence, the improvement offered by an approximate NM over a traditional monitor decreases as the number of actions increases. However, the approximate NM still offer on average a $56\%$ performance improvement over a traditional monitor. When the number of actions is very high (e.g., when the number of actions is 128 in small scenarios), then action preconditions become very complex and most of the times the \textsf{NOP} action is executed by all agents, which entails that the all monitors obtain a good performance. We can see that, as in the previous experiment, the approximate NM performs slightly worse than the full NM (i.e., the full NM offers on average a $15\%$ performance improvement over an approximate NM). However, full and approximate NMs noticeably outperform the traditional monitor regardless of the number of actions.

\begin{table}[h]
\center{
\scriptsize{\hspace{-3pt}
\begin{tabular}{ |c|c|c| }
\hline
\multirow{2}{*}{Actions}&Traditional&Approximate NM\\
&Monitor&Identify+Discover\\
\hline
2&	52\%&		95+8\%\\ \hline
8&	49\%&		75+1\%\\ \hline
32&	32\%&		48+0\%\\ \hline
128&	29\%&		40+0\%\\ \hline
\multicolumn{3}{c}{$G\in\llbracket 1,500\rrbracket$}\\
\end{tabular}
\begin{tabular}{ |c|c|c|c| }
\hline
\multirow{2}{*}{Actions}&Traditional&Full&Approximate NM\\
&Monitor&NM&Identify+Discover\\
\hline
2&	69\%&	99\%&	95+3\%\\ \hline
8&	47\%&	75\%&	71+0\%\\ \hline
32&	38\%&	63\%&	48+0\%\\ \hline
128&	53\%&	83\%&	57+0\%\\ \hline
\multicolumn{4}{c}{$G\in\llbracket 1,5\rrbracket$}\\
\end{tabular}}}

\caption{Fulfilments Detected in the Action Possibilities Experiment}
\label{tab:instF}
\end{table}

\begin{table}[h]
\center{
\scriptsize{\hspace{-3pt}
\begin{tabular}{ |c|c|c| }
\hline
\multirow{2}{*}{Actions}&Traditional&Approximate NM\\
&Monitor&Identify+Discover\\
\hline
2&	53\%&		93+5\%\\ \hline
8&	49\%&		76+6\%\\ \hline
32&	35\%&		50+1\%\\ \hline
128&	31\%&		39+0\%\\ \hline
\multicolumn{3}{c}{$G\in\llbracket 1,500\rrbracket$}\\
\end{tabular}
\begin{tabular}{ |c|c|c|c| }
\hline
\multirow{2}{*}{Actions}&Traditional&Full&Approximate NM\\
&Monitor&NM&Identify+Discover\\
\hline
2&	68\%&	99\%&	95+2\%\\ \hline
8&	48\%&	76\%&	72+1\%\\ \hline
32&	38\%&	62\%&	47+0\%\\ \hline
128&	56\%&	85\%&	59+0\%\\ \hline
\multicolumn{4}{c}{$G\in\llbracket 1,5\rrbracket$}\\
\end{tabular}}}

\caption{Violations Detected in the Action Possibilities Experiment}
\label{tab:instV}
\end{table}

\subsection{Summary}
The conclusions of our evaluation are threefold:
\begin{enumerate}
\item Both approximate and full reconstruction processes are more effective (i.e., detect more norm violations and fulfilments) than traditional monitoring approaches regardless of the scenario complexity (i.e., action possibilities and  observability). Both in the case study and in the random experiments our algorithms improved significantly the percentage of violations and fulfilments detected.  
\item Approximate reconstruction is slighting less effective than full reconstruction. In the case study, where a single prohibition norm was monitored; the approximate NM obtained almost the same results as the full NM. In our random experiments, where several prohibition and obligation norms were monitored, the full NM offered an average improvement of a 18\% over an approximate NM.  
\item Approximate reconstruction is scalable with the scenario size (i.e., the number of agents and actions to be monitored). In particular, our experiments demonstrate that the approximate algorithm can be used to monitor a large number of agents (we simulated scenarios with up to 500 agents), actions (we simulated scenarios with up to 128 actions), and norms (we simulated scenarios with up to 128 norms).
\end{enumerate}
\section{Related Work}\label{sec:rw}
Previous work on norms for regulating MAS proposed control mechanisms for norms to have an effective influence on agent behaviours \cite{grossi2007ubi}. These \textit{control} mechanisms are classified into two main categories \cite{grossi2007ubi}: \textit{regimentation} mechanisms, which make the violation of norms impossible; and \textit{enforcement} mechanisms, which are applied after the detection of norm violations and  fulfilments, reacting upon them.

\textit{Regimentation} mechanisms prevent agents from performing forbidden actions (vs. force agents to perform obligatory actions) by mediating access to resources and the communication channel, such as Electronic Institutions (EIs) \cite{IE}. However, the regimentation of all actions is often difficult or impossible. Furthermore, it is sometimes preferable to allow agents to make flexible decisions about norm compliance \cite{castelfranchi2003formalising}. In response to this need, \textit{enforcement} mechanisms were developed. Proposals on the enforcement of norms can be classified according to the entity that monitors whether norms are fulfilled or not. Specifically, norm compliance can be monitored by either agents themselves or the underlying infrastructure may provide monitoring entities.   

Regarding \textit{agent monitoring}, this approach is characterized by the fact that norm violations and fulfilments are monitored by agents that are involved in an interaction \cite{venkatraman1999verifying,daskalopulu}, or other agents that observe an interaction in which they are not directly involved \cite{sen2007emergence,DBLP:journals/aamas/PinninckSS10,6736104}. The main drawback of proposals based on agent monitoring is the fact that norm monitoring  and enforcement must be implemented by agent programmers. 

Regarding \textit{infrastructural monitoring}, several authors proposed developing entities at the infrastructure level that are in charge of both monitoring and enforcing norms. Cardoso \& Oliveira \cite{cardoso2007institutional} proposed an architecture in which the monitoring and enforcement of norms is made by a single institutional entity. This centralized implementation represents a performance limitation when dealing with a considerable number of agents. To address the performance limitation of centralized approaches, distributed mechanisms for an institutional enforcement of norms were proposed in \cite{minsky2000law,gaertner2007distributed,modgil2009framework,criado2013manea}.

All of the aforementioned proposals on monitoring assume that monitors have complete observational capabilities. Exception to these approaches is the recent work of Bulling et al. \cite{bulling2013monitoring} and Alechina et al. \cite{alechina2014norm}. In \cite{bulling2013monitoring}, the partial observability problem is addressed combining different norm monitors to build ideal monitors (i.e., monitors that together are able to detect the violation of a given set of norms). In \cite{alechina2014norm}, the authors propose to synthesise an approximate set of norms that can be monitored given the observational capabilities of a monitor. However, there are circumstances in which norms cannot be modified (e.g., contract and law monitoring) or ideal monitors are expensive and/or not feasible. We take a different approach in which norms and monitors' observation capabilities remain unchanged and monitors reconstruct unobserved actions. 

Our approach is also related to planning, where methods (e.g., POMDPs \cite{kaelbling1998planning}) for choosing optimal actions in partially observable environments have been proposed. A major difference between these proposals and our proposal is that NMs do not perform practical reasoning, i.e., they do not try to optimise or achieve a practical goal. Instead, NMs perform both deductive and abductive reasoning \cite{paul1993approaches} to reason from observed actions to reach a conclusion about the state of the world, and to infer unobserved actions. 
\section{Conclusion}\label{sec:con}

In this paper, we propose information models and algorithms for monitoring norms under partial action observability by reconstructing unobserved actions from observed actions using two different reconstruction processes: full and approximate. Our experiments demonstrate that both reconstruction processes detect more norm violations than traditional monitoring approaches. Approximate reconstruction performs slightly worse than full reconstruction, whereas its computational cost is much cheaper, making it suitable to be applied in practice. 

The reconstruction algorithms proposed in this paper can be applied to several domains that require action monitoring; from normative MAS \cite{normsAgents}, to intrusion detection systems \cite{hu2008adaboost}, to control systems \cite{6870439} and to intelligent surveillance systems \cite{huang2011multitarget}. As future work, we plan to investigate domain-dependent approximations that could speed up action reconstruction even further.

\bibliographystyle{abbrv}
\bibliography{references}
\end{document}